\documentclass[
 apl, reprint,
 amsmath,amssymb,
]{revtex4-2}

\usepackage{graphicx}
\usepackage{dcolumn}
\usepackage{bm}

\begin{document}


\title{\textbf{Integrated Multi-Wavelength Photonic Architectures for Future Scalable Trapped Ion Quantum Devices} 
}%

\author{Alto Osada}
 \email{Contact author: osada.alto.qiqb@osaka-u.ac.jp}
\author{Koichiro Miyanishi}%
\affiliation{%
Center for Quantum Information and Quantum Biology (QIQB), The University of Osaka, 1-2 Machikaneyama, Toyonaka 560-0043, Japan.
}%

\date{\today}

\begin{abstract}
Recent advances of quantum technologies rely on precise control and integration of quantum objects, and technological breakthrough is anticipated for further scaling up to realize practical applications.  Trapped-ion quantum technology is one of the promising candidates to realize them, while its scalability depends on the development of intra-node scaling up, reproducibility of quantum nodes and photonic interconnection among them. Utilization of integrated photonics instead of free-space optics is a crucial step toward mass production of trapped-ion quantum nodes and manifests itself as useful for laser delivery for various quantum operations and photon detection.  However,
whole architecture of the scalable photonic circuits for them is left unexplored.  In this work, we discuss photonic architectures for trapped-ion quantum devices, in which lasers of multiple wavelengths are delivered to multiple trapping zones within a single chip.  We analyze two methods of configuring nanophotonic waveguides and compare them in terms of loss of total laser power.  This work opens up a new landscape of photonic architecture for quantum technologies not limited to trapped-ion quantum devices.
\end{abstract}

\maketitle


\section{Introduction}

Scaling up a number of controllable objects is a ubiquitous proposition in various technologies, which applies to quantum technologies as well in the scope of practical quantum computation~\cite{Saffman2016-sn, Bruzewicz2019-ac, Kjaergaard2020-er, OBrien2009-ko} and quantum communication~\cite{Gisin2007-ey, Kimble2008-bc}. Among a number of physical implementations of quantum technologies, those using lasers for their quantum operations require a vast amount of optical components, including free-space optics, to be scalable and reproducible.  A promising approach is to replace them by integrated photonics~\cite{Silverstone2016-ue, Soref2006-ou, Wang2019-el} which enables to contain various kind of nanometer-scale optical components within a single centimeter-scale chip to manipulate lasers fed from optical fibers.  

\begin{figure}[t]
\centering
\includegraphics[width=8.6cm]{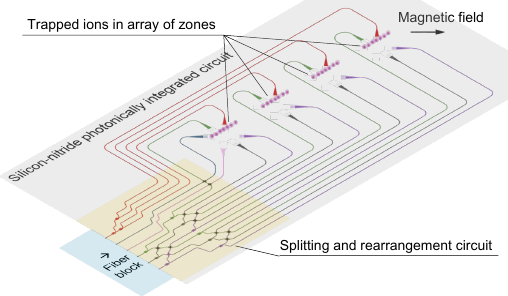}
\caption{Schematic image of a photonic chip that is integrated in trapped-ion quantum device. In the gray-shaded area corresponding to the silicon-nitride based photonically integrated circuit, the area shaded in yellow indicates where the waveguides are split and rearranged toward each zone. The waveguides of different colors, carrying different wavelengths of lasers, reach the designated zone and lasers impinge trapped ions { through electro-optical modulators that modulate and turn on/off the lasers}. Trapping electrodes on the chip are not shown for visibility.}
\label{Fig1}
\end{figure}

Trapped-ion quantum system~\cite{Bruzewicz2019-ac, Stute2012-gz, Leibfried2003-yt, Takahashi2020-fq, Stephenson2020-op, Egan2021-zo, Saha2024-wg, Molmer1999-dq, Wang2021-jp, Harty2014-fm, Kielpinski2002-ot, Krutyanskiy2023-uw, Duan2010-ge} is a good representative, in which lasers of multiple wavelengths ranging from {near-ultraviolet} to near-infrared regions should be frequency-stabilized to a reference, intensity-modulated and delivered at places where ions float inside a vacuum chamber.  So far, laser delivery~\cite{Mehta2016-nu, Niffenegger2020-kb, Mehta2020-yr, Hogle2023-pu, Mordini2024-nb, Kwon2024-zz, Ivory2021-kh} and photon detection~\cite{Todaro2021-sy} have been separately demonstrated with photonic elements integrated just beneath electrodes to apply voltages for trapping ions.  These successful experiments strengthen the idea of quantum CCD architecture~\cite{Kielpinski2002-ot}, where ions are shuttled around above the chip from one trapping zone to another, experiencing quantum gates, measurement, state initialization and so on with the desired lasers at each trapping zone being carried via nanophotonic waveguides.

At this point, a question arises: how are the lasers coming from outside the chip to the waveguides at one of the zones, and how scalable can we make the integrated photonic chip with such functionality?  A straightforward answer to feed lasers for all waveguides is that each waveguide is connected to an optical fiber dedicated to it, resulting in the number of optical fibers sticking to the chip being $mn$, with $m$ being the number of the different colors of lasers and $n$ being the number of zones to deliver the lasers. This can be elaborated when the number of zones and the required lasers are small, however, a more efficient way is anticipated when $n$ becomes on the order of 10.  In such a circumstance, integrated photonics manifests itself as a powerful platform by providing splitters (see e.g. Refs~\cite{Krutov2020-ci, Song2024-zm, Dai2012-wg, Wang2019-el} and references therein) for dividing optical powers as is done in Kwon \textit{et al.}~\cite{Kwon2024-zz}, waveguide crossings (see e.g. Refs~\cite{Nevlacsil2020-mz, Liu2004-wr, Sommer2023-kr, Johnson2020-ss}) for the rearrangement of optical waveguides, and modulators~\cite{Hogle2023-pu} for modulation and switching. A thorough picture of how the photonic chip should be is still missing, and in particular an efficient way to feed every zone with multiple lasers starting with waveguides for only $m$ different wavelengths is desired.

In this work, we propose an architecture of a photonic integrated circuit which enables $m$ optical fibers to feed $mn$ waveguides that may cover full functionality of $n$ trapping zones.  The key is how to split and rearrange the $m$ waveguides into $mn$ as efficient as possible, for which we consider two methods: bubble sort and blockwise duplication we term here.  We compare these two methods in terms of the number of photonic elements, namely the splitters and waveguide crossings, and total power transmission as well.  We also take realistic parameters into account in our estimation to find that, {although the superior method depends on the loss budget of the circuit,} the blockwise duplication method is more {efficient and} scalable {for a set of feasible parameters}.  Our work provides a good starting point to develop a whole architecture of integrated photonics not only for ion trap but also for other material systems that should be scaled up within a single chip.

\section{Results}

\subsection{Multi-wavelength, multi-zone architecture}

We consider a situation that our integrated photonics address $n$ zones on the chip, each of which {contains lasers of $m$ different wavelengths}.  This is actually the case for trapped-ion systems, as exemplified in the following.

It is inevitable to use multiple wavelengths of lasers in trapped-ion quantum technology. In the case of strontium ion, six lasers with wavelengths of 405~nm, 461~nm, 422~nm, 1092~nm, 674~nm, 1033~nm are needed.  The first two wavelengths are transitions of neutral strontium atom and play a role of photoionization~\cite{Mende1995-dj}.  The remaining four are those for $^2S_{1/2}\rightarrow {^2}P_{1/2}$, $^2D_{3/2}\rightarrow {^2}P_{1/2}$,$^2S_{1/2}\rightarrow {^2}D_{5/2}$ and $^2D_{5/2}\rightarrow {^2}P_{3/2}$, respectively.
They are, for instance, utilized for state preparation~\cite{Lin2013-yi, Harty2014-fm}, laser cooling (see e.g. Ref.~\cite{Itano1995-xg}), quantum gates~\cite{Takahashi2017-pu, Haljan2005-my, Schafer2018-fz, Ballance2016-wt, Gaebler2016-rd, Leibfried2003-ny, Roos2008-mw, Harty2014-fm} and quantum state measurement~\cite{Harty2014-fm}.  One to three lasers out of above six are used for each of the quantum operations. Their frequencies are precisely tuned and polarizations and wavevectors are defined so that the selection rules are well satisfied to drive transitions in a desired way.  Practical configurations are provided later in Sec.~IID.

From above example, we can see that a trapping zone has multiple optical outcouplers, like waveguide gratings, dedicated for different wavelengths and they are aligned in a fixed configuration with respect to the direction of magnetic field that defines the quantization axis.  This configuration repeatedly appears for every zone under the assumption of a uniform, static magnetic field.  The situation is depicted in Fig.~\ref{Fig1}.

Given such a situation, our photonic circuit should have the following elements.
First, $m$ spot size converters~\cite{Marchetti2019-nb, Zhang2024-bv, Maegami2015-dr, Brunetti2023-cx} or couplers of any other type receive lasers of different $m$ wavelengths from $m$ optical fibers.  Second, a bunch of splitters and waveguide crossings divides and rearranges the waveguides so that they are arranged in the way the basic unit of $m$ lasers of different wavelengths is repeated $n$ times, $n$ being the number of zones.
Third, we want to turn on/off the laser in every single waveguide independently of each other. This can be done by inserting an electro-optic or any other type of modulator and forming a Mach-Zehnder interferometer in every waveguide just after the rearrangement.  Then finally, the waveguides and the outcouplers at their very end bring the lasers to the zones and irradiate them on targets.  In terms of the complexity and power efficiency of the circuit, the first, third, and final steps are very important in the element-level developments but have no impact on how to configure optical elements.  The second point, on the other hand, is worth consideration, since there should be a lot of splitters and crossings imposing non-zero insertion losses, and the total number may vary depending on the delivery algorithm to yield different power efficiency which determines the scalability of the circuit.  Therefore, we focus on the split-and-rearrangement part in the following, to better develop an efficient photonic architecture.

\begin{figure}[t]
\centering
\includegraphics[width=8.6cm]{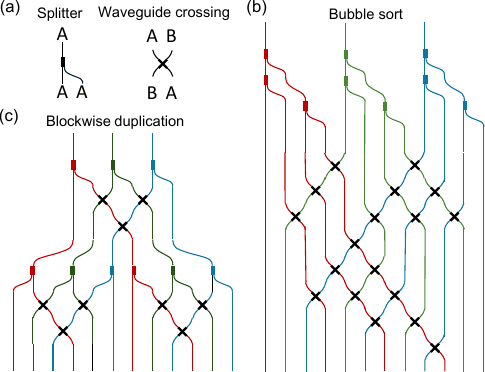}
\caption{(a) Schematics of a splitter and a waveguide crossing. (b), (c) Construction of photonic circuits for rearrangement in the case of $m=3$ and $n=4$ using (b) bubble sort and (c) blockwise duplication, respectively.}
\label{Fig2}
\end{figure}

\begin{table}[b]
\caption{Numbers of photonic elements for $m$ wavelengths and $n$ zones. $N_Y$ stands for the total number of splitters, $n_Y$ the maximum number of splitters for a single waveguide, $N_X$ and $n_X$ those for waveguide crossings.
}
  \label{tab1}
  \centering
\begin{tabular}{ccc}
\hline
Methods & Bubble sort & Blockwise duplication \\
\hline
$N_Y$ & $m(n-1)$ & $m(n-1)$ \\
$n_Y$ & $\sim \log_2{n}$ & $\sim {n}/{2}$ \\
$N_X$ & ${m(m-1)n(n-1)}/{4}$ & $m(m-1)n/2$ \\
$n_X$ & $(m-1)(n-1)$ & $(m-1)\cdot\lceil{n/2}\rceil$ \\
\hline
\end{tabular}
\end{table}

\begin{figure}[t]
\centering
\includegraphics[width=7cm]{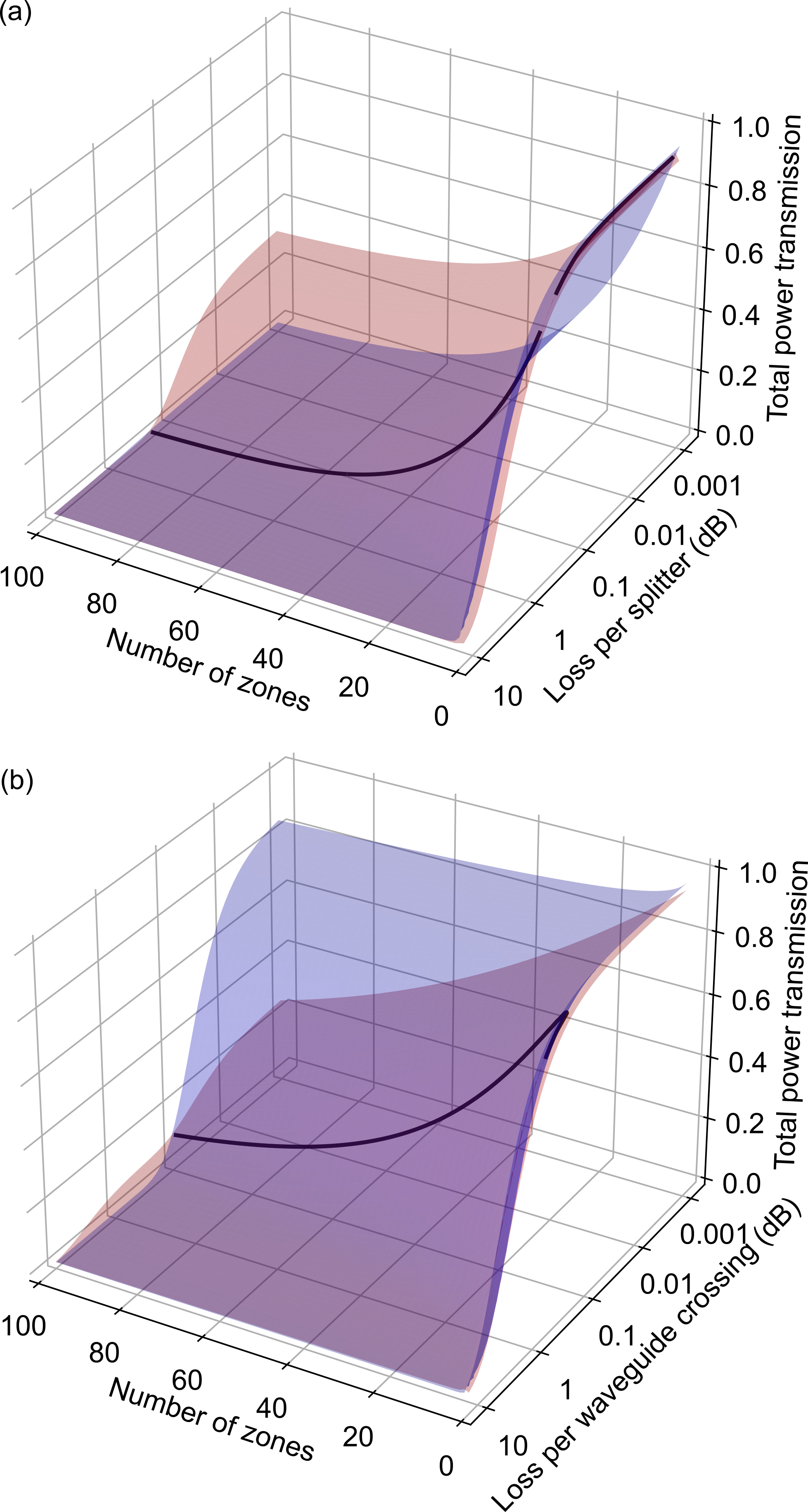}
\caption{Surface plots of the total power transmissions $T_{\rm{block}}$ (red) and $T_{\rm{bubble}}$ (blue) for $m=3$, (a) $\eta_X = -0.22$~dB and $\eta_Y$ varied and (b) $\eta_Y = -0.1$~dB and $\eta_X$ varied. Black curves indicate the intersections of the red and blue surfaces {as guide to the eye.}}
\label{Fig3}
\end{figure}

\subsection{Comparison of the methods}

Here we consider two methods, which we call bubble sort and blockwise duplication.   The first one, bubble sort, is to split each of the $m$ waveguides into $n$, namely, we have $mn$ waveguides, and apply bubble sort to it to rearrange as desired.  The other method, which we call blockwise duplication, is to split each waveguide of $m$ wavelengths into two first, apply swaps next to rearrange it to have $m$ colors of lasers bundled together in two blocks, and then repeatedly duplicate the block on both sides of the sequence until it coincides with the desired arrangement.  The problem gets simpler by regarding $m$ waveguides at the beginning as $m$ letters and manipulating the sequence of letters to make a target sequence, see Appendix A for details.

We compare these methods in several properties. The numbers of splitters and waveguide crossings are important indicators of how simple the circuit is. The maximum numbers of splitters and crossings experienced by a single waveguide are also of interest in practical perspectives, namely in terms of the design of the power splitters. The numbers themselves are of course important, however, we are interested in the scaling of them with respect to $m$ and $n$, which tells us which method is better scaled up.  We try to figure out, for various parameters, which method is better by setting a figure of merit later.

\begin{figure}[t]
\centering
\includegraphics[width=8.6cm]{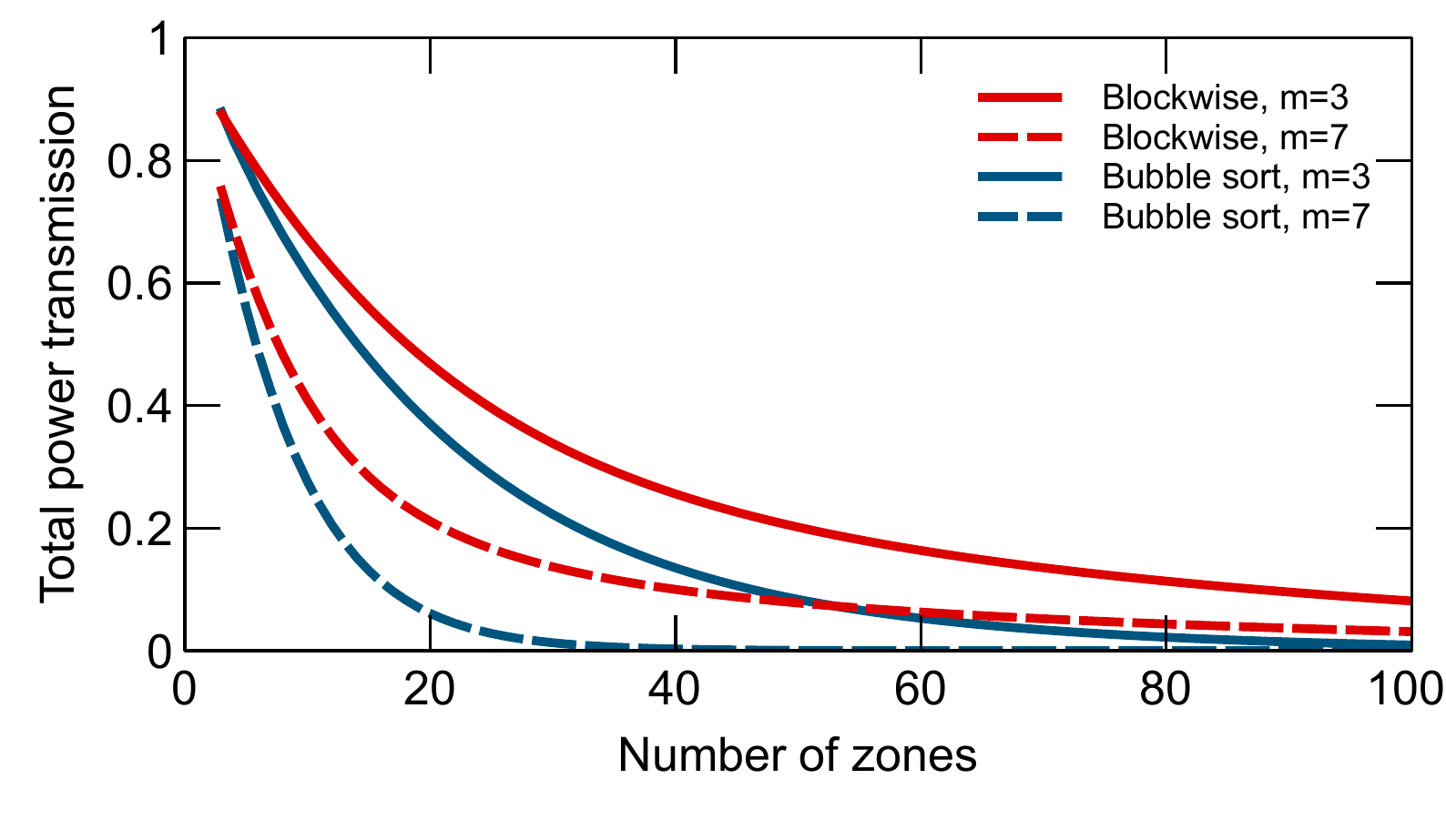}
\caption{Total power transmissions $T_{\rm{block}}$ (red) and $T_{\rm{bubble}}$ (blue) for $m=3$ (solid) and $m=7$ (dashed) with $\eta_X = -0.22$~dB and $\eta_Y = -0.1$~dB.}
\label{Fig4}
\end{figure}

The bubble sort method seems to be the most straightforward way of solving this game of waveguide rearrangement.  Since each splitter adds one waveguide, the total number of splitters is $m(n-1)$. The maximal number of splitters that a single waveguide undergoes scales as $\log_2{n}$, which is very slow compared to $n$.  The total number of crossings is calculated as $m(m-1)n(n-1)/4$.  The maximum number of crossings a single waveguide undergoes is $(m-1)(n-1)$, since the waveguide with rightmost wavelength should cross $(n-1)$ copies and $(m-1)$ colors of waveguides to get to the leftmost block.  

The blockwise duplication method allows the number of waveguide crossings to be much smaller and scales slower as well, as we shall see in the following.  This method is modular in the sense that we can calculate for a single duplication procedure and add up $n$ times.  The first duplication includes $m$ splitters and $m(m-1)/2$ crossings.  Following procedures will take place always on the leftmost block to add on the left or the rightmost block to add on the right, therefore, the total numbers of splitters and crossings increases as $nm(m-1)/2$.  Note that the copies of the blocks can be generated on both sides of the letter sequence, the maximum number of splits for a single letter scales as $n/2$.  The maximum number of crossing that a single waveguide may pass is $(m-1)(n-1)$, which is the same number as the bubble sort. 

\begin{figure*}[t]
\centering
\includegraphics[width=16cm]{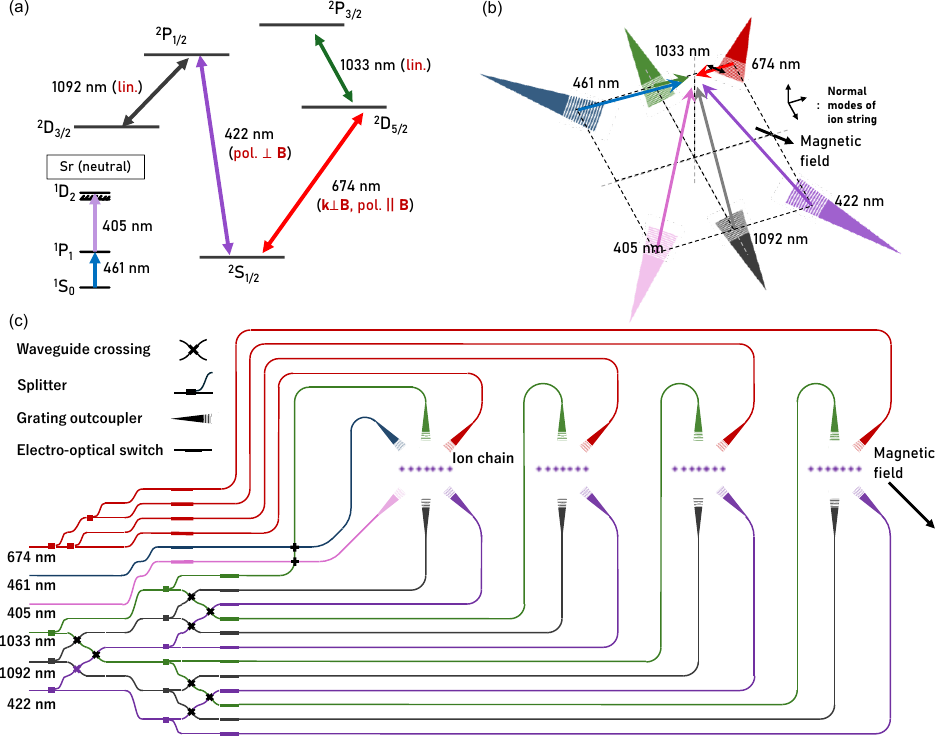}
\caption{(a) Simplified energy level structure of strontium ion and neutral strontium atom.  {Polarization ("pol.") and wavevector ($\mathbf{k}$) are specified here in red, where "lin." refers to the linear polarization in any possible axis, "pol. $\perp \mathbf{B}$" to the polarization perpendicular to the magnetic field, and so on.}  (b) Configuration of six lasers required in a single zone with an applied magnetic field indicated by a thick black arrow.  A black arrow attached to the laser direction represents the linear polarization of the laser field. (c) Possible configuration of a photonic circuit in a chip delivering six lasers to four trapping zones.  The waveguides (solid lines) of the same color belong to the same wavelength. Photonic elements are indicated in the upper left  of the figure. }
\label{Fig5}
\end{figure*}

These are summarized in Table~\ref{tab1}.  The splitter and waveguide crossing are schematically depicted in Fig.~\ref{Fig2}(a).  Figures~\ref{Fig2}(b) and ~\ref{Fig2}(c) show examples of the implementations of the two methods for the case of $m=3$ and $n=4$. One might readily see the modularity of blockwise duplication, while bubble sort lacks it, which results in the smaller circuit size of the blockwise duplication method.  Blockwise duplication is advantageous not only in the circuit size but also in the scaling of the total number of crossings, however, it has a disadvantage in { the maximum number of splitters that a single waveguide encounters}, in which the bubble sort method does well.

{
Let us compare the possible sizes, or area occupied by, the circuit made by these methods.  Since waveguides are as thin as 1~$\mu$m or less and are bent at will with radii of curvatures larger than the minimum bend radii, we can ignore their sizes and focus on the size of other photonic elements, i.e., the splitters and waveguide crossings.  By taking the largest elements into account, we can regard the sizes of those elements as the same for all wavelengths.  Then, the circuit sizes are simply determined by the number of elements, leading to the conclusion that a circuit built by the blockwise duplication method occupies smaller area, for the better scaling of the total number of waveguide crossings.
}
Realistic estimation of maximal number of zones with given laser powers with loss budgets will be discussed in Sec.~IID.

\subsection{Comparison by total power transmission}

To quantitatively compare these methods in depth, we define a total power transmission $T$ that sums up the power output from all waveguides of the same wavelength, normalized by the input power and averaged over $m$ wavelengths, to use it as a figure of merit of the method.  We assume that all waveguides are provided with the same power, all splitters and crossings have the same transmission coefficients $\eta_Y$ and $\eta_X$, respectively, and the waveguides are lossless.  Counting the numbers of factors $\eta_X$ and $\eta_Y$ for the two methods (see Appendices A and B), the total power transmissions are calculated and plotted in Figs.~\ref{Fig3} and \ref{Fig4}. {Here we approximate the numbers of splitters that a single waveguide undergoes as $log_2{n}$.}  Figure~\ref{Fig3}(a) displays the total power transmission for bubble sort $T_{\rm{bubble}}$ (blue) and blockwise duplication $T_{\rm{block}}$ (red) as functions of number of zones and transmission per a single splitter with {$\eta_X = -0.22$~dB, which is obtained separately by our numerical calculation to be published elsewhere}.  With a small loss of a splitter up to about {0.1~dB}, the blockwise duplication overwhelms the bubble sort thanks to the better scaling of the number of crossings with respect to $n$. 

The total power transmissions for the two methods are shown in Fig.~\ref{Fig3}(b) when the number of zones and the transmission per waveguide crossing are varied and $\eta_Y = -0.1$~dB fixed as a feasible value~\cite{Krutov2020-ci}.  With the loss per waveguide crossing being smaller than about {0.1~dB}, the bubble sort becomes better since the poor scaling of the number of crossings has less impact and better scaling property of the maximal number of splitters that a single waveguide undergoes benefits more.  As the loss per waveguide crossing increases, the plots for the bubble sort rapidly fall as expected, and blockwise duplication becomes the least bad method.

For values {$\eta_X = -0.22$~dB} and $\eta_Y = -0.1$~dB~\cite{Krutov2020-ci}, the $T_{\rm{block}}$ (red) and $T_{\rm{bubble}}$ (blue) are plotted in Fig.~\ref{Fig4}.  The solid and dashed lines, respectively, show those for $m=3$ and $m=7$, supposing the trapped-ion quantum devices containing single species of ion or two species of ions while they may vary depending on the roles and the species of ions.  We see, from the displayed results, that the curve of the blockwise duplication method falls slower than that of the bubble sort as $m$ and $n$ increase, which infers that the blockwise duplication scales no worse than the bubble sort.  {One might even consider the blockwise duplication method as a better way for its attractive aspects such as the modularity and the small circuit size.}

\subsection{Design for strontium ions}

As a demonstration of how our architecture will be implemented on a photonic circuit, we will provide a concrete example of implementation for strontium ions with the number of zones being four.  While we consider strontium ions here, the configuration given here is directly applicable for ions with similar energy structures, e.g. Ca$^+$ and Ba$^+$ simply by replacing the values of wavelengths.  

Figure~\ref{Fig5}(a) shows the energy structure related to trapped-ion quantum operations and ion generation, from which we see that out of six wavelengths, 405~nm and 461~nm lasers are for photoionization loading of ions and not required in every zone.  {Polarization ("pol.") and wavevector ($\mathbf{k}$) are specified here in red, where "lin." refers to the linear polarization in any possible axis, "pol. $\perp \mathbf{B}$" to the polarization perpendicular to the magnetic field, and so on.  }  If we adopt grating outcouplers, the configuration depicted in Fig.~\ref{Fig5}(b) can cover the required quantum operations.  {Note that the lasers are assumed to have linear polarizations that is allowed by the transverse electric mode of the waveguide gratings.}   Here we also assume the quantum CCD architecture~\cite{Kielpinski2002-ot} that allows ions to move around from one zone to another, {or within the zone to address different ions in an ion chain}.  Therefore, it is sufficient to deliver 405~nm and 461~nm lasers to a certain zone.  The remaining four lasers of 422~nm, 1092~nm, 674~nm and 1033~nm wavelength are to be delivered to every zone.  Here we notice that by accessing from top and the bottom, or equivalently from left or right, the four lasers can be grouped in two and splitting and rearrangement can be done within the groups.  If we assign one wavelength to the first group and remaining wavelengths to the second, there is no need for the waveguide crossing in the first group, minimizing the loss of that laser.  We here assign 674~nm for this, which requires much more power than others to speed up quantum operations.  {Another thing to note is that another, or more, waveguide gratings might be needed for 674~nm-wavelength laser depending on the applications e.g., two-qubit quantum gates and/or addressing multiple ions in a zone simultaneously without shuttling.}

As a result, we can draw a photonic circuit for the situation we consider here as in Fig.~\ref{Fig5}(c).  We did not consider the bend radius of the waveguides properly in this figure; however, for silicon nitride for the wavelengths considered here, the bend radius can be about a few hundreds of $\mu$m, which makes this circuit easily contained within a realistic chip for ion trapping. {The splitting-and-rearrangement part is configured by the blockwise duplication method as an example.}

\begin{table}[b]
\caption{
Estimated powers required at a single zone and typical output powers of single, commercial diode lasers.
}
  \label{tab2}
  \centering
\begin{tabular}{ccc}
\hline
Wavelength (nm) & Req. power (mW) & Typ. output (mW) \\
\hline
405 & $> 0.1$ & 10 \\
461 & $> 0.1$ & 200 \\
422 & $> 0.1$ & 50 \\
1092 & $> 0.1$ & 50 \\
674   & $> 10$  & 300 \\
1033   & $> 0.1$ & 50 \\
\hline
\end{tabular}
\end{table}

\begin{figure}[t]
\centering
\includegraphics[width=8.6cm]{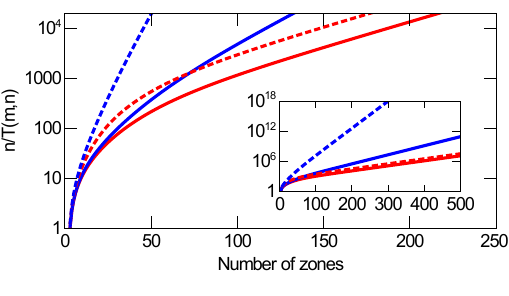}
\caption{Right hand side of the inequality $P_{\rm{laser}}/P_{\rm{out}} > n/T(m,n)$ which determines the available number of zones with sufficient laser power.  Red and blue curves are for blockwise duplication and bubble sort, respectively. Solid curves represent the results for $m=3$ and dotted ones those for $m=7$.}
\label{Fig6}
\end{figure}

We characterize our photonic architecture for this specific case further in terms of the maximal number of trapping zones that can be supported within a single chip equipped with one optical fiber for each wavelength of the laser.  
Estimated powers of lasers required at each zone are $100$~$\mu$W for all lasers except for the 674~nm one for which 10~mW is desirable to ensure reasonable speed of quantum operations, and then the output powers of the single commercial lasers matter.  These are listed in Table.~\ref{tab2} and we see that the limiting factor is the 674~nm laser, whose typical output power affords 30 zones at most, calculated from the listed parameters.  Assuming that a single zone holds $n_i$ ions with $n_i$ being in the range from 1 to 10, a single chip can support about several tens to a few hundreds of ions that are immediately addressed by lasers in such a situation.  The limitation by the 674~nm laser will be overcome if we simply add more lasers and corresponding fibers generating and carrying 674~nm lasers. If we loosen the initial proposition to allow such a situation, the limitation is imposed by other lasers of wavelengths such as 422~nm/1092~nm/1033~nm. Under this condition, the number of accessible zones and ions can be increased to a few hundreds of zones and thousands of ions, with the best case that the laser power is fully available for the irradiation to ions.  For more quantitative estimation of the number of available zones, we use the total power transmission $T(m,n)$, input laser power $P_{\rm{laser}}$ and laser power at the output coupler $P_{\rm{out}}$ to obtain the condition that the lasers can be supplied to every zone as
\begin{align*}
    \frac{P_{\rm{laser}}T(m,n)}{n} \ge P_{\rm{out}}
\end{align*}
which is rewritten as $P_{\rm{laser}}/P_{\rm{out}} \ge n/T(m,n)$.  The right hand side of the inequality is plotted in Fig.~\ref{Fig6}, where red and blue curves are for blockwise duplication and bubble sort, respectively. Solid curves represent the results for $m=3$ and dotted ones those for $m=7$.  The value of the $P_{\rm{laser}}/P_{\rm{out}}$, which is purely given by experiments, is set first and compared with $n/T(m,n)$ to see the maximally available number of zones.  For example, if we consider 422~nm laser with Table.~II kept in mind, we have $P_{\rm{laser}}/P_{\rm{out}} = 500$ and for $m=3$ we see from Fig.~\ref{Fig6} that 72 zones can be fed with sufficient laser power.  With this number of zones, a single trap chip can support several hundreds of ions and larger $P_{\rm{laser}}$ and smaller $P_{\rm{out}}$ will make the number of ions, or identically the number of qubits, larger.

\section{Discussion}

{First, it should be noted that the photonic architecture considered here is not solely for a one-dimensional array of trapping zones, but is also applicable to a two-dimensional array.  For a two-dimensional array forming an $M\times N$ square lattice, for example, $M$ trapping zones in a column are regarded as a large block to form an array of $N$ blocks.  This is possible when $(M-1)m$ waveguides can pass through between the trapping zones, separated by 1~mm at least.  $M$ and $N$ will then be at most on the order of ten, given that the waveguides are bundled with their spacings of $\sim $10~$\mu$m.}

In the previous Section, we saw that the blockwise duplication method is better for {a feasible set of} parameters.  However, there is a drawback in terms of the design of the splitters.  The bubble sort method, splitting the waveguide and hence the power at first steps, can be ideally constructed by only $m$ designs of the 50:50 splitter for $m$ wavelengths in the case that $n$ is some power of 2.  In blockwise duplication, it is by nature required that $\sim n/2\times m$ different splitters are designed.  However, considering the number of crossings each waveguide undergoes, different designs of splitters of about $m\log_2{n}$ are needed for the bubble sort method, as inferred in Appendix A.

One can consider an intermediate method between bubble sort and blockwise duplication, namely one first splits the waveguide into $\nu$ ($\nu = 3,4,\dots$) and makes $\nu$ blocks of waveguides, which will be repeated until the number of blocks reaches $n$.  Depending on the loss budget, such a method may yield better total power transmission than the two methods described in this paper, which will be addressed in future works.

Another issue is that the photonic circuit architecture we provided possesses the waveguide crossings of different wavelengths at its heart, so that we should design $m(m-1)/2$ different bicolor waveguide crossings with their losses made very small.  We will make a proposal on this issue elsewhere.

We consider other ways to deliver the waveguides.  Indeed, we can deliver the waveguides without waveguide crossings, if we can manage the $\sim mn$ waveguides inside the area just beneath the trapping zones, just as exemplified in Appendix C.  We did not consider this possibility for the following reasons: first, the trapping zones are a few tens to a couple of hundred micrometers in size, which is by nature comparable to the height of the ions off the chip.  Assuming the bend radii of the silicon nitride waveguides are as small as 100~$\mu$m for safety, but not too large, it is still hard to manage all of them inside the trapping zones.  Next, the waveguides should pass in between the grating outcouplers, which are several tens of $\mu$m, which limits the number of waveguides that can pass through the gap, hindering scaling up.  The scheme may also have a disadvantage in the maximal waveguide length, where  waveguides should go through every zone to deliver waveguides to the last zone, in contrast to the methods using waveguide crossing that can avoid such a situation.  This is particularly harmful when the waveguide loss is not negligible {as in the present case where the propagation loss of $> 1$~dB/cm is inevitable for wavelengths shorter than 500~nm~\cite{Corato-Zanarella2024-fo}}. An example of a laser delivery circuit is shown in Appendix C.

The possibility of using photonic circuits consisting of multiple layers of silicon nitride and/or other material platforms should also be mentioned.  As we see in Fig.~\ref{Fig5}, two wavelengths per single layer can be delivered to the zones without waveguide crossings.  Therefore, there is a possibility of fabricating multilayered photonic circuits including $\sim m/2$ layers.  This idea faces the difficulties of coupling lasers from optical fibers to the waveguides since the use of the standard fiber array is not facile, and the processing and alignments become harder { with every elements in the photonic circuit being kept high quality}.  

\section{Conclusion}

In this work, we considered a photonic chip architecture with multiple wavelengths addressing multiple zones in the chip.  Starting from the consideration of required photonic elements, we proposed to prepare an optical fiber carrying laser of each wavelength, split and rearrange the waveguides inside the chip, and deliver them to the designated zones instead of attaching a vast amount of optical fibers equal to the total number of waveguides at the zones.  We compared two methods for building such a split-and-rearrangement circuit, the bubble sort and blockwise duplication.  {For a set of realistic parameters, blockwise duplication showed better performance.  For scaling up, this method will be useful for its smaller circuit size and modularity, facilitating the practical design of the photonic chip.}  We mainly focused on the use of the architecture for trapped-ion quantum technology; however, it is also applicable to other quantum and classical technologies requiring the delivery of multiple wavelengths in a broad range to multiple zones.  To our knowledge, our work provides the first quantitative analysis of such a photonic architecture for quantum devices in terms of scalability and feasibility, broadening the scopes of quantum and photonic technologies.

This work was supported by JST Moonshot R\&D program (JPMJMS2063-5-2).





\bibliography{main}




\appendix

\section{Derivation of Table.~1}

\begin{figure}[b]
\centering
\includegraphics[width=7cm]{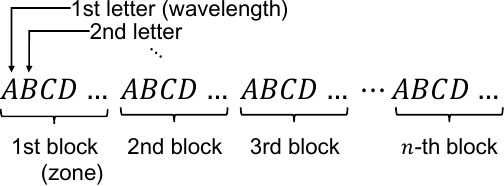}
\caption{The target sequence where $m$ letters forming a block, which is repeated $n$ times.}
\label{FigS1}
\end{figure}

To properly deliver $m$ lasers to the designated zones, the splitters and crossings are connected to realize the desired configuration, and the problem is the algorithm to determine the circuit.  There are two allowed elements, splitter and crossing, and we map this problem as follows. First, we have a sequence of $m$ different letters $\underbrace{ABCD\dots}$ where different letters corresponds to wavguides of different wavelengths.  Using a \textit{split} $A\rightarrow AA$ and \textit{swap} $AB \rightarrow BA$ operations, we want to achieve a target sequence $\underbrace{ABCD\dots} \underbrace{ABCD\dots} \, \cdots \, \underbrace{ABCD\dots}$ consisting of $mn$ letters. Figure~\ref{FigS1} shows the target sequence with the correspondence between letters and wavelengths, and blocks and zones.  Here we consider two methods, which we call bubble sort and blockwise duplication in the main text.   The first one, bubble sort, is to split each of the $m$ letters into $n$, namely, we have $AA\dots BB\dots CC\dots DD \dots \,\cdots\,$ which has $mn$ letters, and apply bubble sort to it to have the target sequence.  The other method, which we call blockwise duplication, is to duplicate each letter once first to make $AABBCCDD\dots$, apply swaps next to rearrange it as $\underbrace{ABCD\dots} \underbrace{ABCD\dots}$, and then repeatedly duplicate the block $\underbrace{ABCD\dots}$ on both sides of the sequence until it coincides with the target sequence.


The bubble sort method seems to be the most straightforward way of solving this game of sequence making.  Since each split adds one letter to the sequence, the total number of splits reads $m(n-1)$. The maximal number of splits that a single letter undergoes scales as $\mathrm{log}_2{n}$, which is slow compared to $n$.  The total number of swaps is calculated as follows. For the sequence $AA\dots BB\dots CC\dots DD \dots \,\cdots\, $, we apply $n-1$ swaps to bring the leftmost $B$ letter to the left to make $ABAA\dots BB\dots CC\dots DD \dots \,\cdots\,$.  In this manner, $ABAB\dots ABCC\dots DD\dots$ is realized by $(n-1)+(n-2)+\cdots+1 = n(n-1)/2$ swaps.  In the same way, bringing every $i$-th letter to the left to bring at the rightmost position of the block $ABC\dots$ takes $in(n-1)/2$, where $i$ runs from $1$ to $m-1$.  Therefore, the total number of swaps reads $m(m-1)n(n-1)/4$.  The maximum number of swaps a single letter undergoes is $(m-1)(n-1)$.

The blockwise duplication method allows the number of waveguide crossings to be much smaller and scales slower as well, as we shall see in the following.  This method is modular in the sense that we can calculate for a single duplication procedure and add up $n$ times.  The first duplication goes from $\underbrace{ABCD\dots} \rightarrow AABBCCDD\dots \rightarrow \underbrace{ABCD\dots} \underbrace{ABCD\dots}$.  In the second transformation, $m(m-1)/2$ swaps are executed.  The number of swaps the letters undergo is $0, 1, 2, \cdots,(m-2),  (m-1), (m-1), (m-2), (m-3), \cdots , 2, 1, 0$ from the left.  As a next step, we would like to duplicate the right block to add a new block on the right, by additional $m$ splits and $m(m-1)/2$ swaps, and the number of swaps the letters undergo is $0, 1, 2, \cdots, (m-1)$ for the left block, $(m-1), (m-1), \cdots, (m-1)$ for the block in the middle, and $2(m-1), 2(m-2), 2(m-3), \cdots , 4, 2, 0$ for the newly added block on the right.  Following procedures will take place always on the leftmost block to add on the left or the rightmost block to add on the right, therefore, the total numbers of splits and swaps increases as $n$ gets large, and the maximum number of swaps a letter undergoes grows as $(m-1)(n-1)$.  Note that copies of blocks can be generated on both sides of the letter sequence, the maximum number of splits for a single letter scales as $n/2$.

\begin{figure}[t]
\centering
\includegraphics[width=7cm]{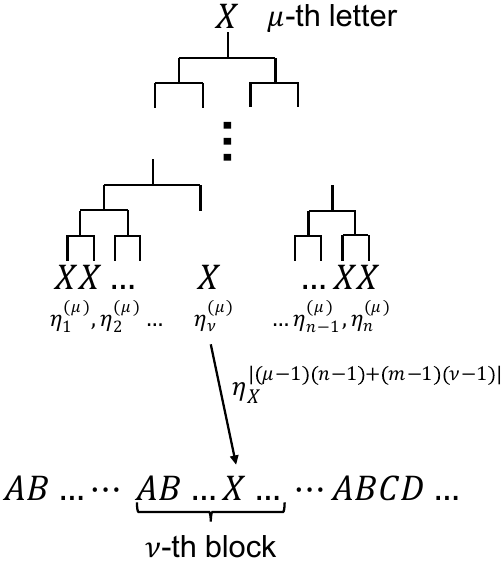}
\caption{Schematic representation of the bubble sort method to bring $\mu$-th letter $X$ to $\mu$-th position of $\nu$-th block.  }
\label{FigS2}
\end{figure}

\begin{figure}[t]
\centering
\includegraphics[width=7cm]{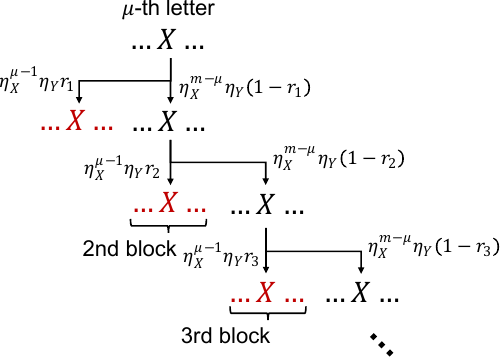}
\caption{Schematic representation of the blockwise duplication method.  Blocks in red manifests themselves as having no more split and swap.}
\label{FigS3}
\end{figure}

\section{Calculation of total power transmission}

To assess which method analyzed above is better in a more quantitative way, we consider the total power transmission $T_{\rm{bubble}}$ and $T_{\rm{block}}$ for each method, measuring the ratio of total power transmitted through the $n$ waveguides of the same wavelength to the input laser power averaged over $m$ wavelengths.  To make these quantities meaningful, practical figures of merit, we assume that $n$ waveguides of the same wavelength should carry the same optical power, which complicates the calculations.  We shall calculate $T_{\rm{bubble}}$ and $T_{\rm{block}}$ in the following.

\subsection{Bubble sort}

We consider a $\mu$-th letter out of $m$, and how it is delivered to $\nu$-th block out of $n$, as depicted in Fig.~\ref{FigS2}  We should first define the transmission just after the splitters $\eta_{\nu}^{(\mu)}$ for the $\mu$-th letter going to $\nu$-th block. These are $1/n$ if splits and swaps are lossless, though it is not the case in reality. For unideal transmissions of swaps $\eta_X$ and splits $\eta_Y$, branching ratios of the splitters varies.  For the $\mu$-th letter to arrive at $\mu$-th position of $\nu$-th block, it experiences { $|(n-1)(\mu-1)-(m-1)(\nu-1)|$ swaps.  Therefore, the transmission coefficient $t_{\nu}^{(\mu)}$ for each waveguide is }
\begin{align}
    t_{\nu}^{(\mu)} = \eta_Y^{\log_2{n}} {\eta_X^{|(n-1)(\mu-1)-(m-1)(\nu-1)|}} \eta_{\nu}^{(\mu)} \label{S1}
\end{align}
and these takes the same value for $\nu = 1, 2, \dots, n$ from the assumption that the optical power is equally divided to $n$.  We see that 
\begin{align}
  {  \eta_{\nu+1}^{(\mu)} = \frac{\eta_X^{|(n-1)(\mu-1)-(m-1)(\nu-1)|}}{\eta_X^{|(n-1)(\mu-1)-(m-1)\nu|}} \eta_{\nu}^{(\mu)} }
\end{align}
{for $\nu = 1, 2, \dots, n-1$.  These recurrence equations enable us to calculate Eq.~\ref{S1} and thus}
\begin{align}
    T_{\rm{bubble}} = \frac{1}{m} \sum_{\mu = 1}^{m} \sum_{\nu = 1}^{n} {t_{\nu}^{(\mu)}}  = \frac{1}{m} \sum_{\mu = 1}^{m} n {t^{(\mu)}}.
\end{align}
Here we put $t^{(\mu)} = t_{\nu}^{(\mu)}$ relevant for all $\mu =1, 2, \dots, m$.

\subsection{Blockwise duplication}

In blockwise duplication, a $\mu$-th letter is split and swapped to the next block, and the split and swapped over and over until the target sequence is formed. The splitters should be designed to have all different branching ratios by nature.  The procedures are depicted in Fig.~\ref{FigS3}.  The branching ratio of the splitter to make $(\nu+1)$-th copy is determined to be $(1-r^{(\mu)}_\nu):r^{(\mu)}_\nu$, where we define it so that $r^{(\mu)}_\nu$ goes to the left and $(1-r^{(\mu)}_\nu)$ to the right.  Each time one duplicates a block, the $\mu$-th letter on the right should undergo $m-\mu$ crossings to the right and the $\mu$-th letter on the left is crossed by $\mu-1$ letters.  The added factors throughout these procedures are indicated in the vicinity of the arrows in Fig.~\ref{FigS3}.  We get a transmission coefficient $t_\nu^{(\mu)}$ as 
\begin{align}
    t_\nu^{(\mu)} =
    \begin{cases}
    \eta_Y^{\nu} \eta_X^{(\nu-1)(m-\mu)+(\mu-1)} r^{(\mu)}_\nu \prod_{\nu' = 1}^{\nu-1} (1-r^{(\mu)}_{\nu'}) & \text{for $\nu = 2, 3, \dots, n-1$,} \\
    \eta_Y^{n} \eta_X^{(n-1)(m-\mu)}\prod_{\nu' = 1}^{n-1} (1-r^{(\mu)}_{\nu'}) & \text{for $\nu = n$} .
  \end{cases}
\end{align}
These quantities are assumed to be equal, that is, 
\begin{align}
    t_2^{(\mu)} = t_3^{(\mu)} = \cdots = t_n^{(\mu)}. \label{S5}
\end{align}
The first block was not included here because it is totally symmetrically duplicated over and over to the left to have the same sequence of the same set of transmission coefficients, if we split the power equally by setting $r^{(\mu)}_1 = \eta_X^{m-1}/(1+\eta_X^{m-1})$.  Equations~(\ref{S5}) yields a set of recurrence equations for $r^{(\mu)}_\nu$ expressed as
\begin{align}
    r^{(\mu)}_{n-1} &= \frac{\eta_X^{m-\mu}}{\eta_X^{m-\mu} + \eta_X^{\mu -1}}, \\
    r^{(\mu)}_{\nu} &= \frac{\eta_Y \eta_X^{m-\mu} r^{(\mu)}_{\nu+1}}{1+\eta_Y \eta_X^{m-\mu} r^{(\mu)}_{\nu+1}} \hspace{5mm} \text{for $\nu = 2, 3, \dots, n-2$} ,
\end{align}
from which we get $ r_\nu^{(\mu)}$'s and thus $ t_\nu^{(\mu)}$'s to further calculate the total power transmission as 
\begin{align}
    T_{\rm{block}} = \frac{2}{m} \sum_{\mu = 1}^{m} \sum_{\nu = 2}^{n} {t_{\nu}^{(\mu)}}  = \frac{1}{m} \sum_{\mu = 1}^{m} 2(n-1) t_{\nu}^{(\mu)}.
\end{align}
The factor of $2(n-1)$ reflects that $2(n-1)$ blocks are generated through the duplication procedure shown in Fig.~\ref{FigS3}, if the procedures are executed symmetrically to the left and right.

\begin{figure*}[t]
\centering
\includegraphics[width=16cm]{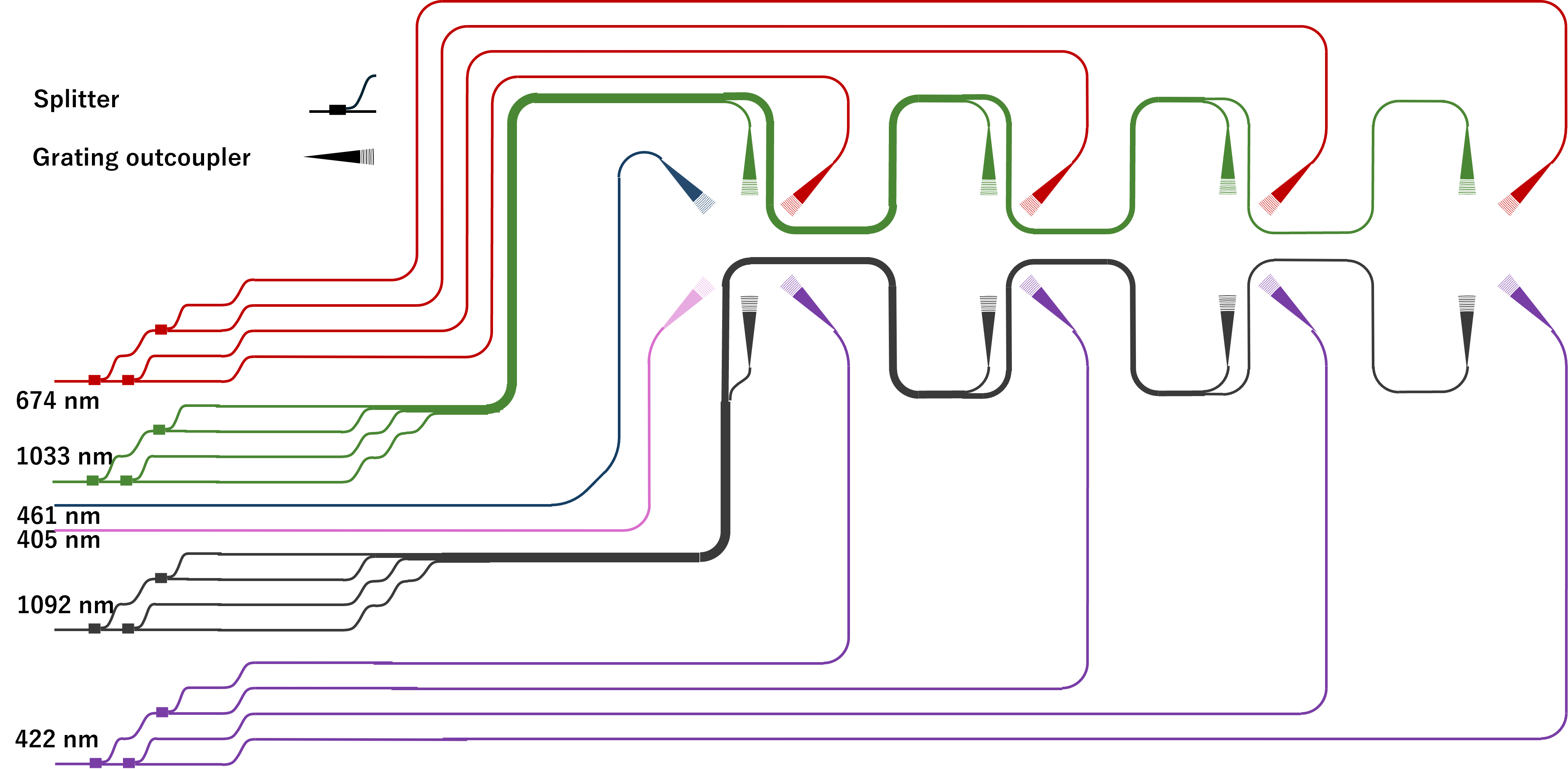}
\caption{Schematics of how to deliver lasers without waveguide crossings, for the case of $n=4$.  Here the bending radii and the width of the bundles of waveguides do not reflect the realistic ones.}
\label{FigS4}
\end{figure*}

\section{Photonic circuit without crossing}

Figure ~\ref{FigS4} schematically displays an example of the photonic circuit delivering lasers to four trapping zones, as in Fig.~5(c) in the main text.  This method delivers the lasers in one-dimensional manner, that is, the longest waveguides of a certain wavelength should transit all other zones till the destination.  We did not consider this way of constructing laser delivery circuit as valid, however, it might be a good choice if the following conditions are satisfied, even though the ultimate scalability is not equipped with it.  First, the gratings should be placed several hundreds of $\mu$m away from the lateral trapping position of the ions, which allows a bunch of waveguides can pass between the grating outcouplers.  Second, and more importantly, the waveguide losses are sufficiently small so that the transmitted power surpasses those of the bubble sort and blockwise duplication methods.  Since trapped-ion systems inevitably involve more visible wavelengths than near-infrared wavelengths, this method seems to be not realistic.

\end{document}